\documentstyle[prl,aps,amsfonts,amssymb,twocolumn,epsfig]{revtex}

\begin{document}
%
 \twocolumn[\hsize\textwidth\columnwidth\hsize\csname @twocolumnfalse\endcsname

\title{Giant mass and anomalous mobility of particles in a fermionic system}
\author{Achim Rosch and Thilo Kopp}
\address{Institut f\"ur Theorie der Kondensierten Materie, 
Universit\"at Karlsruhe,
D--76128 Karlsruhe, Germany}

\date{\today}
\maketitle

\begin{abstract}
  We calculate the mobility of a heavy particle coupled to a Fermi sea
  within a non-perturbative approach valid at all temperatures. The
  interplay of particle recoil and of strong coupling effects, leading
  to the ortho\-go\-na\-lity catastrophe for an infinitely heavy
  particle, is carefully taken into account.  We find two novel types
  of strong coupling effects: a new low energy scale $T^{\star}$ and a giant
  mass renormalization in the case of either near-resonant scattering
  or a large transport cross section $\sigma$.  The mobility is shown
  to obey two different power laws below and above $T^{\star}$.  For
  $\sigma\gg\lambda_f^2$, where $\lambda_f$ is the Fermi wave length,
  an exponentially large effective mass suppresses the mobility.
\end{abstract}

\vskip1.5pc]

Some remarkable advances in condensed matter physics are based on a
thorough analysis of models of a single impurity coupled to a
many-particle system.  The polaron problem and the mobility of ions in
liquid $^3$He are well-known examples for the dynamical behavior of a
delocalized particle in a bosonic or fermionic bath, respectively.  We
will consider the latter case of a delocalized but heavy particle
strongly interacting with a Fermi liquid.

Concerning the experimental context, a $^3$He$^{+}$-ion in
normal-fluid $^3$He is the cleanest possible heavy-particle
Fermi-liquid system. Unfortunately, the transition to the superfluid
state cuts off the available low-temperature range \cite{woelfle}.
The particle is indeed heavy and of large cross section since the ion
forms a snowball: about 100 Helium atoms (depending on pressure) are
tightly bound at temperatures $T$ up to several hundred mK.  Other
real fermionic systems, such as muons in metals or valence band holes
in $n$-type doped semiconductors, show a more complex dynamical
behavior, usually involving several diffusion mechanisms
\cite{hartmann,kagan,kondo,yama}. Nevertheless, our approach is supposed
to be valid in the low-$T$ regime.
 
We will discuss two strong coupling effects of a heavy
particle\cite{IntDegrees,PhDthesis} in a fermionic bath, both of which
have not been reported before, basically because a microscopic strong
coupling calculation of the mobility has not been developed for low
temperatures. As to the origin of strong coupling effects we actually
have to distinguish two scenarios: (1) the scattering phase at the
Fermi energy of bath particles is close to $\delta=\pi/2$ in any
single scattering channel; (2) the transport cross section $\sigma$ of
the heavy particle is large: $\sigma \gg (2\pi/k_f)^2$. Obviously,
each of the two scenarios implies a strong coupling of bath particles
to the impurity.

In the first scenario we work close to a resonance at the Fermi energy.
This may signify that an attractive channel exists with sufficiently
large binding energy.  Alternatively, we may consider a strong
enough repulsive interaction between the particle and the fermionic
bath, e.g.\ the Hubbard model in the spin-polarized sector with one
overturned spin which interacts with the free Fermi system of up-spins
through a local interaction $U$. If the tight-binding band is nearly
half-filled with up-spins and the repulsive interaction approaches
infinity, $U\rightarrow\infty$, the bath particles at the Fermi energy
encounter a phase shift close to $\pi/2$. These problems
cannot be solved within perturbation theory. Indeed,
the heavy-particle mobility changes its temperature
dependence drastically with respect to the perturbational limit if the
phase shift exceeds a minimal value in at least one channel: the
temperature (and frequency) power law behavior acquires an anomalous
exponent as we will demonstrate.

The second scenario may be encountered already for relatively small
phase shifts in each of the scattering channels provided the particle
couples to the bath through a sufficiently large number $N$ of
channels. We find that the effective mass of the particle is
tremendously increased for low $T$: it scales with an exponential
function, the exponent of which is proportional to a power of $N$.

The mobility of a heavy particle in a fermionic environment has been
investigated for more than three decades
\cite{woelfle,josephson,kondo,yama}.  Concerning translational
invariant systems, as ions in $^3$He, the work by W\"olfle et al.\ 
summarized the previous findings and presented the first unified
treatment for all temperature ranges below the Fermi temperature
within a self-consistent Mori-Zwanzig scheme. The approach is
phenomenological and cannot account for part of the strong coupling
effects discussed here. More recently Prokof'ev\cite{prokof} put
forward a microscopic path integral scheme, yet his evaluation is
restricted to the high temperature corrections in the mobility.

\begin{figure}[h] 
\epsfig{width=0.95 \linewidth,file=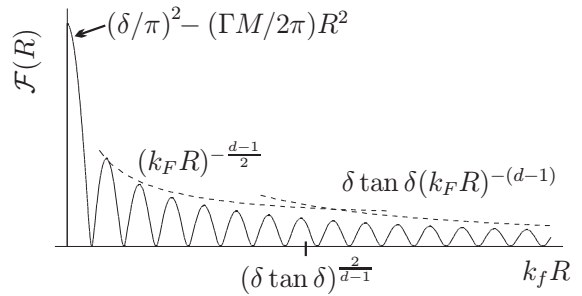}
\caption{Qualitative behavior of ${\cal F}(R)$ for strong coupling.}
\label{fig1}
\end{figure}

The basis of our evaluation (for space dimension $d>1$) is an
effective low-energy action for the heavy particle, first introduced
by Sols and Guinea \cite{sols}, investigated in more detail by
Prokof'ev\cite{prokof}, and evaluated for the heavy-particle spectral
function in a previous paper\cite{rosch}:
\begin{equation}
S_{\text{eff}}=S_o  - {1\over2}\int\!\!\!
\int^\beta_0 \!\!\! d\tau 
d\tau'{\cal F}(|\bbox{R}(\tau')\!-\!\bbox{R}(\tau)|)\,
\rho(\tau'\!\!-\!\tau).
\label{Seff}
\end{equation}
The first term on the rhs is the action for a free particle, $
S_o\!=\!{1\over2}\int_0^\beta\!d\tau M_o \dot {\bbox{R}}(\tau)^2 $,
and ${\cal F}(|\bbox{R'}-\bbox{R}|)$ is determined through the overlap
of the electronic ground state $\phi_{\bbox{R}}$ with particle at
position $\bbox{R}$ and a state $\phi_{\bbox{R'}}$ in the
thermodynamic limit $N_e\rightarrow\infty$ \cite{yamada}: $
\langle\phi_{\bbox{R'}}|\phi_{\bbox{R}}\rangle\rightarrow
  \exp\{-[(\delta/\pi)^2-{\cal F}(|\bbox{R'}-\bbox{R}|)] \log N_e\} 
  $ where
\begin{equation}{\cal F}(|\bbox{R}|)=
    (\delta/\pi)^2-{1\over\pi^2}\arcsin^2(\sqrt{1-x^2} \sin\delta)
\label{F}\end{equation}
with $x\!=\!\langle e^{i\bbox{kR}}\rangle_{|\bbox{k}|=k_f}$ (see
Fig.~\ref{fig1}).  These relations apply for $s$-wave scattering with
phase shift $\delta$.  Generalizations for several scattering channels
are found in the literature \cite{yamada,vladar}. The leading
low-energy form of $\rho(\tau\!-\tau')$ in $d>1$ \cite{comment1} is
$\rho(\tau\!-\tau') =({\pi/\beta})\sum_{n\neq 0} |\omega_n| \,
e^{i\omega_n (\tau-\tau')}$ which reflects that the number of virtual
excitations close to the Fermi surface is proportional to $\omega$.
This behavior and the Friedel oscillations of ${\cal F}(R)$ are rooted
in the fermionic character of the bath.

Alternatively, one could set up a Hamiltonian for both, the heavy
particle (${\bbox{P}}$,${\bbox{R}}$) and the fermionic degrees of
freedom $(c^{\dagger}_{\bbox{k}},c_{\bbox{k}})$
\begin{equation}
H={\bbox{P}^{\,2}\over 2M_o} + 
\sum\nolimits_{\bbox{k}} {k^{\,2}\over 2m} \,c^{\dag}_{\bbox{k}} 
c_{\bbox{k}} + 
 U\sum\nolimits_{\bbox{kk'}} e^{i\,(\bbox{k'}-\bbox{k})\bbox{R}}
    \; c^{\dag}_{\bbox{k}} c_{\bbox{k'}}
\label{H}\end{equation} for $s$-wave scattering.
In a second step, the fermions are traced out. This procedure results
in an effective heavy-particle action which involves any number of
retardations (i.e.\ time integrals) and it cannot be handled further
except for a few limiting cases. Fortunately, the effective
action~(\ref{Seff}) reproduces all exactly known limits of this more
basic approach: the spectral function of the heavy particle was
evaluated and discussed in this respect \cite{rosch}, and the mobility
$\mu(T)$ is identical in both approaches in the exactly known
high-temperature limit: a constant behavior
$\mu(T)\!=\!\mu_o\!=\!e/(M_o \Gamma)$ plus logarithmic corrections
\cite{prokof,rosch}. Here $\Gamma\!=\!n k_f\sigma/M_o$ is the
scattering rate, $k_f\!=\!2\pi/\lambda_f$ the Fermi wave vector and
$n$ the particle density.

The success of the approach based on the effective action~(\ref{Seff})
is actually not surprising. It correctly accounts for the two
competing effects which control the dynamics: the orthogonality
catastrophe (OC) through the exact form of the overlap and the recoil
through translational invariance. The relevant low-energy scale is
supposed to settle which of the two effects dominates. It may be
anticipated that the maximum recoil energy $E_{R}= (2k_f)^2/2M_o$
transferred in a collision process is the relevant energy scale. For
frequency $\omega$ or temperature $T$ far above $E_{R}$ the energy
balance in scattering processes is independent of the recoil---i.e.\ 
physics is governed by the $\bbox{R}=0$ saddle point and the
corresponding fluctuations: the high-$T$ mobility results from the
coefficient of the $R^2$-term in ${\cal F}(R)=(\delta/\pi)^2 - (\Gamma
M_o/2\pi)\, R^2 + O(R^4)$.  
For frequencies or temperatures much less than $E_{R}$ the available
phase space for scattering processes with momentum transfer
$|\Delta\bbox{k}|\simeq 2k_f$ is restricted by recoil, i.e.\ in $d$
dimensions by a factor $(T/E_{R})^{(d-1)/2}$ and we expect a
well-defined quasi particle for $T\rightarrow 0$ ---albeit with
renormalized mass or possibly coupling constant. For the mobility this
corresponds to a $1/T^{(d+1)/2}$ power law which is the Fermi liquid
behavior well known for many decades from weak coupling analysis
\cite{woelfle}. It is instructive to realize that the decay of the
envelope of the overlap function ${\cal F}(R)$ for $k_f R\gg1$
determines the low temperature behavior: firstly, only a decaying
envelope function results in a finite quasiparticle weight $Z$
\cite{rosch}, and secondly, the long-range decay
$\propto\delta\tan\delta/(k_f R)^{d-1}$ translates into the power-law
behavior of the incoherent part of the spectral function and the
mobility.

Here we want to investigate the significance of the second energy
scale, the particle-bath interaction $U$. The cross section of the
particle is directly related to the interaction via the phase shift
$\delta$: $n k_f\sigma=M_o \Gamma=E_{R}\sin^2\delta/\pi d$.  In
phenomenological approaches it was always explicitly or implicitly
assumed that the interaction enters $\mu(T)$ only through the cross
section. However the phase shift may also lead to a subtle
interference effect for the moving particle. This interference effect
is encoded in the spatial decay of ${\cal F}(R)$ (cf.\ 
Fig.~\ref{fig1}). It decays anomalously slow for
$\delta\rightarrow\pi/2$: ${\cal F}(R)\propto 1/(k_f R)^{(d-1)/2}$ for
$1\ll k_f R\ll (\delta\tan\delta)^{2/(d-1)}\equiv k_f R_c$ whereas for
larger $R$ the envelope function decays with $1/(k_f R)^{d-1}$, as
stated above. For $\delta=\pi/2$, the anomalous decay extends to
$R\rightarrow\infty$. The cross-over length $R_c$ corresponds to a new
low-energy scale $E^{\star}=E_{R}/(\delta\tan\delta)^{{4/(d-1)}}$
which approaches $E^{\star}\rightarrow 0$ with $\delta\rightarrow
\pi/2$ and which exists only for $\tan\delta \gg 1$, i.e.\ we consider
a strong-coupling anomaly. This observation implies that physical
quantities will behave differently for $T$ (or $\omega$) below or
above $E^{\star}$. Specifically, the power-law behavior above
$E^{\star}$ is characterized by an anomalous exponent ---as we will
show for $\mu(T)$.  A similar anomalous enhancement of interference
effects near $\delta\simeq\pi/2$ may be relevant for other fermionic
problems with dynamical scatterers at distinct spatial positions.

\begin{figure}[h]
\epsfig{width=0.95 \linewidth,file=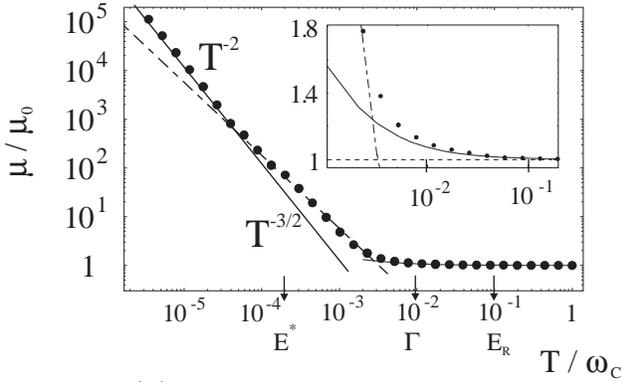}
\vspace{0.0truecm}
\caption{$\mu(T)$ in the strong coupling
  regime with $\delta=0.95\pi/2$, $\omega_c\simeq \epsilon_f$. The
  inset displays $\mu(T)$ on a linear scale together with the
  analytical result of the high-$T$ expansion.}
\label{fig2}
\end{figure}
We evaluated the mobility in the simplest scheme available which still
reproduces the exact high $T$ behavior, including the non-perturbative
logarithmic corrections. It also reproduces the Born-approximation
results of the previous literature which are commonly agreed upon as
to display the correct power-law behavior for the lowest temperature
regime. Self-consistency is indispensable for a successful evaluation
since a diffusing particle exhibits an entirely different long-time
behavior as compared to a free particle. In order to tackle the
mobility $\mu_{ij}(\omega)=-(1/i\omega)\,\langle\dot{R}_i
\dot{R}_j\rangle_{ (\omega+i0)}$ we have to examine a $\langle R
R\rangle_S$--correlator which is a path integral evaluated with action
$S$ (Eq.~(\ref{Seff})). For this purpose we introduce a general
quadratic action
\begin{equation}
S_{\text{V}}[A]=S_o  - {1\over4}\int\!\!\!
\int^\beta_0 \!\!\! d\tau d\tau' A(\tau\!\!-\!\tau')\,
\Bigl(\bbox{R}(\tau)\!-\!\bbox{R}(\tau')\Bigr)^2
\label{SO}
\end{equation}
and expand the correlator to first order in $\Delta S=
S-S_{\text{V}}$: $\langle R R\rangle_S = \langle R
R\rangle_{S_{\text{V}}} - \langle (R R) \Delta S\rangle_{S_{\text{V}}}
+ \langle R R\rangle_{S_{\text{V}}} \langle\Delta
S\rangle_{S_{\text{V}}} +O(\Delta S^2)$. The best possible action
$S_{\text{V}}[A]$ to this order results from the condition that only
the first term survives in next-to-leading order, i.e.\ $A(\tau)$ has
to be found from $ \langle (R R) \Delta S\rangle_{S_{\text{V}}} =
\langle R R\rangle_{S_{\text{V}}} \langle\Delta
S\rangle_{S_{\text{V}}}
$.
And the mobility is
\begin{equation}
\mu(\omega+i0,T)=\biggl[{A(\omega+i0,T)\over i\omega} - i\omega M_o\biggr]^{-1}
\label{mobility}
\end{equation}
where $A(\omega+i0,T)$ is the analytical continuation of the Fourier
transform of $A(\tau)$.  It is straightforward to show that this
procedure is equivalent to Feynman's variational principal (FVP). The
early FVP evaluations of the mobility in the polaron problem did not
fully reproduce the weak coupling limit. However only few variational
parameters were used to determine $A(\tau)$. These inconsistencies are
repaired if the full function $A(\tau)$ is determined
self-consistently. Details of this calculation will be presented in a
more extended article.  Fig.~\ref{fig2} displays the calculated the
static $\mu(T)$ in $d=3$. The phase shift is close enough to $\pi/2$
so that the anomalous behavior $\mu(T)\approx T^{-3/2}$ extends over
two decades.

The second pronounced strong coupling effect, the giant mass
renormalization, is illustrated most intuitively via a rough estimate
based on a Kramers-Kronig relation. It relates mass $M(\omega)$ and
the friction coefficient $\eta(\omega)$ as defined by the exact
inverse mobility (at $T=0$): $\mu(\omega+i0)^{-1}=\eta(\omega)-i
\omega M(\omega)$. The friction $\eta(\omega)$ \cite{comment3} attains
its high energy value $\eta'$ for frequencies larger than the relevant
energy scale $E_{R}$ (and below $\epsilon_f$). For $\omega\ll E_{R}$
the phase space for scattering processes approaches 0 so that
$\eta(\omega)$ converges algebraically to zero.  The
zero-frequency $T=0$ mass enhancement is therefore estimated by a
step-like $\eta(\omega)$: $\Delta M \propto \int^{\epsilon_f}_{E_{R}}
d\omega \eta(\omega)/\omega^2 \simeq \eta'/E_{R}=
(\eta'/E^{(o)}_{R})\cdot (M_o+\Delta M)/M_o$. The index $o$ refers to
unrenormalized quantities. Since the friction coefficient is
proportional to the cross section of the particle, or rather
$\eta'/E^{(o)}_{R}\simeq M_o (k_f a)^2$ where $a$ is the diameter of the
particle, we find from this estimate that $\Delta M/M_o\sim
(a/\lambda_f)^2/[1- (a/\lambda_f)^2]$ which diverges for
$a\sim\lambda_f$. While the divergence is an artifact of this estimate
it nevertheless signals that the large number of virtual low energy
excitations for $\omega\simeq E_R \ll \epsilon_f$ enhances the
dynamical mass very effectively.

\begin{figure}[h]
\epsfig{width=0.95 \linewidth,file=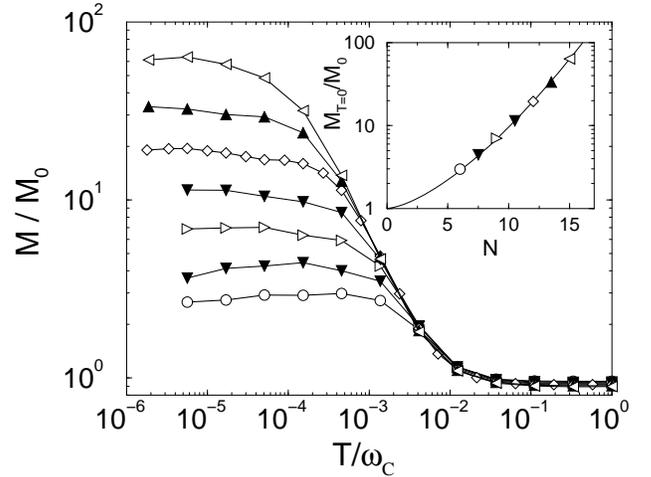}
\caption{Effective mass $M(T)$ for $M_o=40 m$ and $\delta=0.95\pi/2$.
  The respective number of scattering channels is found by comparison
  with the inset which displays the low-$T$ effective mass (the number
  of channels ranges from 6 to 15).}
\label{fig3}
\end{figure}
The $T$-dependent dynamical mass has to be calculated through the
self-consistent scheme outlined above. We have to extend it for a large
number of scattering channels in order to raise $\sigma$ far beyond
$\lambda_f^2$. One may include more scattering channels by systematically
generalizing ${\cal F}(R)$ as has been accomplished by Vladar \cite{vladar}
who included higher angular momentum channels $N$ up to $N=3$.

\begin{figure}[h]
\epsfig{width=0.95 \linewidth,file=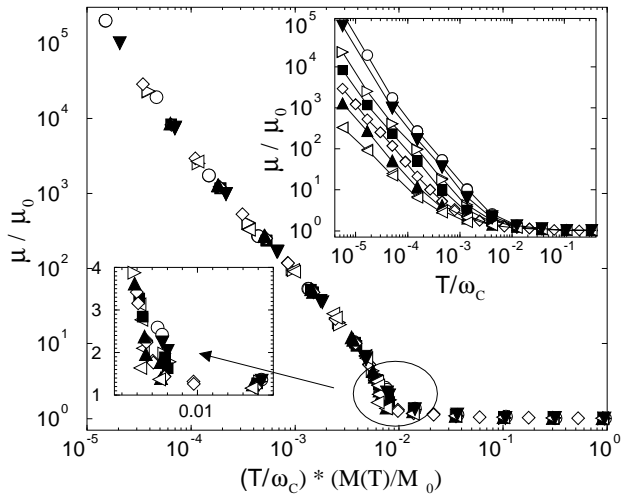}
\caption{Scaled mobility of the $N$-channel model for 
  the same parameters as in Fig.~3. The upper inset displays the
  unscaled mobility (the highest mobility corresponds to 6 channels,
  the lowest to 15 channels implying a large cross section).  The
  lower inset demonstrates that scaling fails for intermediate
  temperatures.}
\label{fig4}
\end{figure}

However we do not intend to model a specific scatterer. In this case,
it is more practical to introduce a large number of spin channels to
investigate this type of strong coupling effect in a generic way. The
number of scattering channels then enters as a multiplicative factor
in the exponent of the overlap: ${\cal F}_N (R) \equiv N {\cal F}
(R)$. Accordingly, the inverse high-$T$ mobility scales with a trivial
factor $N$: $\mu^{-1}_{o_N}=N \mu^{-1}_{o}$ ---and in leading order of
perturbation
theory (${\cal F}_N (R)\ll 1$) this trivial dependence on $N$ is valid
down to $T\rightarrow0$. The self-consistency, however results in a
tremendous suppression of the low-$T$ mobility (Fig.~\ref{fig4}) and a
giant increase of the zero-frequency effective mass (Fig.~\ref{fig3}).
The numerical analysis shows that
\begin{equation}
{M_{T\rightarrow 0}\over M_o}
\simeq \exp\biggl[c\,\Bigl({\Gamma\over E^{(o)}_{R}}\Bigr)^{3/2}\biggr],
\quad {\Gamma\over E^{(o)}_{R}}\sim ({a\over\lambda_f})^2\gg1
\label{mass}
\end{equation}
is an excellent fit (Fig.~\ref{fig3}, inset). The factor $c$ depends
weakly on the phase shift and is approximately $c\simeq 0.21$ in the
perturbative regime and $c\simeq 0.68$ for $\delta=0.95\, \pi/2$.

An exponential increase of $M/M_o$ is indeed expected from the
requirement for a smooth transition from the high-$T$ 
to the low-$T$ regime: the high-$T$ analysis yields for $\mu(T)$ a constant
plus logarithmic corrections down to  $T_{s}\simeq \Gamma
\exp(c' \Gamma/E^{(o)}_{R})$ \cite{josephson}. On the other hand, the low-$T$ behavior is
found analytically because self-consistency is not required: 
$\mu(T)^{-1}\simeq \pi^{-2}\delta\tan\delta \cdot M\, T\, (T/E_R)^{(d-1)/2}$.
Assuming that low- and high-$T$ expressions connect smoothly at
$T_{s}$, we have to demand that 
$
{M_{T\rightarrow 0}/ M_o}
\,\geq\, 2 {\Gamma/ T_{s}}\,\approx\,
\exp\Bigl(c' {\Gamma/ E^{(o)}_{R}}\Bigr)
$
This estimate does not yield the same exponent as the full numerical
calculation but it supports the notion of an exponentially large mass.

Also the mobility depends exponentially on the cross section as is displayed 
in the upper inset of Fig.~\ref{fig4}. The low-$T$ mobility, however,
is explained entirely by the mass renormalization. Consequently, 
all curves of the upper inset collapse onto a single line for the 
lowest $T$. This is not true for intermediate temperatures: the lower 
inset magnifies this non-universal regime. 

The dramatic rise in the effective mass of large heavy particles
should be clearly observable for ions in $^3$He. While the superfluid
transition will inhibit the observation of the low-temperature
effective mass, we expect a large increase of the effective mass
$\propto 1/T$ in the regime where a logarithmic increase of the
mobility has been observed \cite{comment2}.  The experimental
situation is more complex for muons in metals and will be discussed
elsewhere. Our predictions are relevant in a regime where the
temperature is low enough so that $\mu(T)>\mu_0$ holds where $\mu_0$
is the mobility in the absence of a periodic potential (which can be
determined from fits of the mobility to the Kondo-Yamada theory
\cite{kondo,yama}). Unfortunately, all mobility experiments for muons
in metals (even in aluminum) seem to be in the opposite regime
$\mu(T)<\mu_0$ where the effect of the periodic potential determines
the mobility.  Therefore one either has to reach temperatures below
10mK in aluminum or find different metals with more mobile muons.

We acknowledge many helpful discussions with \hbox{N.~Andrei}, R.~v.~Baltz,
H.~Castella,
T.~Costi, A.E.~Ruckenstein, and especially P.~W\"olfle.
This work was supported  by the DFG (T.K.).


\begin{references}
\bibitem{IntDegrees}We reduce the
considerations to sufficiently low energies so
   that all internal degrees are frozen.
\bibitem{comment1}an extended form, valid for $d=1$, is
                      found in \cite{rosch}.
\bibitem{comment2}The effective mass must be measured in a low-frequency
   ($\hbar \omega < k_B T$) mobility experiment; see \cite{PhDthesis} for more
   details.
\end{references}
\end{document}